\documentclass{article}
\usepackage{spconf,amsmath,graphicx,hyperref}
\usepackage{booktabs,multirow,makecell,xcolor,amsfonts}
\usepackage{adjustbox}
\usepackage{graphicx}

\usepackage[T1]{fontenc}
\usepackage[utf8]{inputenc}
\usepackage{pifont}
\usepackage{newunicodechar}
\newunicodechar{✘}{\ding{55}}

\title{SLM-TTA: A Framework for Test-Time Adaptation of Generative Spoken Language Models}
\name{
  \begin{tabular}{@{}c@{}}
    Yuan-Kuei Wu$^{1,2}$\sthanks{Work done during internship at Meta.}, Yang Liu$^{2}$, Yiteng Huang$^{2}$, Zhaojun Yang$^{2}$, Haibin Wu$^{2}$, Ruizhe Huang$^{2}$\\Yi-Te(Ethan) Hsu$^{2}$, Shuyu Kong$^{2}$, Ming Sun$^{2}$, Florian Metze$^{2}$, Li Wan$^{2}$
  \end{tabular}
}
\address{$^{1}$Affil One \qquad $^{2}$Affil Two}

\address{$^1$Graduate Institute of Communication Engineering, National Taiwan University, Taiwan \\
$^2$Meta, USA}

\begin{document}
%
\maketitle

\begin{abstract}
Spoken Language Models (SLMs) are increasingly central to modern speech-driven applications, but performance degrades under acoustic shift—real-world noise, reverberation, and microphone variation. Prior solutions rely on offline domain adaptation, which is post-hoc, data-intensive, and slow. We introduce the first test-time adaptation (TTA) framework for generative SLMs that process interleaved audio–text prompts. Our method updates a small, targeted subset of parameters during inference using only the incoming utterance, requiring no source data or labels. This stabilizes token distributions and improves robustness to acoustic variability without degrading core task accuracy. Evaluated on automatic speech recognition, speech translation, and 19 audio understanding tasks from AIR-Bench, our approach yields consistent gains under diverse corruptions. Because adaptation touches only a small fraction of weights, it is both compute- and memory-efficient, supporting deployment on resource-constrained platforms. This work enhances the robustness and adaptability of generative SLMs for real-world speech-driven applications.
\end{abstract}

\section{Introduction}
\label{sec:introduction}
SLMs are emerging as foundational components in modern speech-driven systems, enabling applications ranging from multilingual virtual assistants to hands-free user interfaces by processing both speech and text prompts to generate coherent textual responses. The emergence of powerful models\cite{abouelenin2025phi,chu2024qwen2,tang2023salmonn,wang2023slm,gong2023joint} highlights a significant trend toward unified models that follow complex instructions over long audio contexts. However, a critical vulnerability threatens their real-world deployment: acoustic shift. Real-world variations like background noise, reverberation, and different microphones, which are often underrepresented in training data, can severely degrade generation quality and reliability, leading to frustrating user experiences and eroding trust in these systems.

A primary method for tackling this problem is offline domain adaptation\cite{wang2025self}. This approach involves collecting labeled or unlabeled data from a target environment and fine-tuning the model, often using parameter-efficient methods like LoRA\cite{hu2022lora}. Although effective, this approach is fundamentally reactive and inefficient. It requires a heavy data collection and training pipeline for every new domain, resulting in a collection of static, specialized model checkpoints that cannot adapt to acoustic conditions in real time.

TTA\cite{wang2021tent,niu2022efficient} offers a distinct approach: adapting model parameters during inference using only the incoming test audio, without relying on source data or ground-truth labels. In the speech domain, TTA has largely been developed for ASR encoder-decoder architectures. For example, SUTA\cite{lin22b_interspeech} introduced single-utterance source-free adaptation; SGEM\cite{kim23f_interspeech} applied sequence-level entropy minimization using beam candidates; and LI-TTA\cite{yoon24c_interspeech} incorporated guidance from an external language model during adaptation. However, these techniques are designed for CTC or seq2seq ASR systems and do not extend to generative SLMs that produce outputs from audio and text prompts.

Beyond ASR, TTA has been applied to contrastive audio-language models in retrieval-style setups\cite{10889886}, typically for zero-shot classification using prompts or adapters. Yet, these models are non-generative and do not address challenges like maintaining stable token distributions over extended generations. A related concept, test-time compute (TTC), improves robustness by increasing inference-time computation without updating model weights. TTC has shown gains on auditory reasoning tasks\cite{dang2025scaling}, but by design, it does not adapt to the current acoustic environment.

This paper is the first to address the gap in TTA for generative SLMs that process interleaved audio and text prompts. We propose a novel TTA framework that dynamically adapts a small subset of model parameters during inference to mitigate the effects of acoustic shift, enhancing the robustness of generation without requiring labeled data or prior domain knowledge. Our work introduces the first parameter-updating TTA method tailored to modern generative SLMs, unlike previous efforts that focus on ASR or non-generative settings. We validate our approach across speech recognition, speech translation, and 19 diverse audio understanding tasks in the AIR-Bench benchmark\cite{yang-etal-2024-air}, demonstrating consistent improvements under various acoustic corruptions while preserving instruction-following behavior. Moreover, our method is computationally efficient, as it updates only a small fraction of the model’s weights, making it a practical solution for resource-constrained real-world deployment.

\section{Method}
\label{sec:method}

\subsection{Problem Formulation}

We consider a SLM, denoted as $f_\theta$, which takes as input an audio signal $x \in \mathbb{R}^T$ and a text prompt $u \in \mathcal{V}$, where $\mathcal{V}$ represents the vocabulary. The model generates a sequence of tokens from a learned codebook $\mathcal{Q}$, enabling tasks such as transcription, translation, and other audio-conditioned language generation. This generation process is autoregressive, at each timestep, the model produces a probability distribution over $\mathcal{Q}$ conditioned on the input and previously generated tokens:
\begin{equation}
p_\theta(y_t \mid x, u, y_{<t}) = \mathrm{Softmax}\big(z_\theta(x, u, y_{<t})\big) \in \Delta^{|\mathcal{Q}|}.
\end{equation}

The core challenge we address is domain shift. While the model is trained on clean audio data from a source domain $\mathcal{D}_s$, it is deployed on real-world inputs that are often corrupted by distortions such as noise or reverberation. We represent these degraded inputs as $\tilde{x} = \mathcal{C}(x)$, where $\mathcal{C}$ denotes an unknown corruption process.

To adapt to such shifts at test time, we partition the model parameters $\theta$ into a frozen subset, $\theta_\mathrm{F}$, and an adaptable subset, $\theta_\mathrm{A}$, which includes selected components of the audio encoder and language model. Given a batch of unlabeled, corrupted test samples
\[
\mathcal{B} = \{(\tilde{x}_b, u_b)\}_{b=1}^B,
\]
all from the same task, our objective is to update $\theta_\mathrm{A}$ online using only the test data itself.

For each test sample $\tilde{x}_b$, the model first performs autoregressive decoding to generate a token sequence $\hat{y}_{b,1:T_b}$, along with the stepwise probability distributions:
\begin{equation}
p_{b,t} = p_\theta(\cdot \mid \tilde{x}_b, u_b, \hat{y}_{b,<t}) \in \Delta^{|\mathcal{Q}|}, \quad t = 1, \dots, T_b.
\end{equation}
These predicted sequences and their associated distributions serve as the foundation for our unsupervised adaptation objective.

\begin{figure}[t]
  \centering
  \includegraphics[width=\linewidth]{Figures/SLM_TTA.pdf}
  \caption{Overview of our SLM-TTA framework. Audio passes through an encoder and projector to condition the Spoken Language Model; a tokenizer produces text tokens. At test time, we optimize either an entropy minimization or a pseudo-labeling objective, with the loss gated by a confidence-aware filter.}
  \label{fig:slm_tta}
\end{figure}

\subsection{Episodic Test-Time Adaptation}

Figure~\ref{fig:slm_tta} illustrates the overall structure of our TTA framework, which performs adaptation at test time using a stateless, episodic strategy. Rather than continuously updating the model, we adapt it separately for each incoming batch of test data. Specifically, for each batch $\mathcal{B}$, we begin from the original pre-trained parameters $\theta_0 = (\theta_{\mathrm{A},0}, \theta_{\mathrm{F},0})$.

We first select one of the unsupervised objectives, either entropy minimization or pseudo-labeling, as the loss function for the current batch. We then apply confidence-aware filtering to this loss, retaining only high-confidence tokens for gradient updates. Using this filtered loss, we perform $K$ steps of gradient descent to update the adaptable parameters $\theta_\mathrm{A}$. The updated model is then used to generate predictions for the current batch. Finally, we reset $\theta_\mathrm{A} \leftarrow \theta_{\mathrm{A},0}$ before moving on to the next batch, ensuring that no adaptation state carries over between batches.

The parameter update rule for adaptation step $k$ with learning rate $\eta > 0$ is:
\begin{equation}
    \theta_\mathrm{A}^{(k+1)} = \theta_\mathrm{A}^{(k)} - \eta \nabla_{\theta_\mathrm{A}^{(k)}} \mathcal{L}\big(\mathcal{B}; \theta_\mathrm{A}^{(k)}, \theta_\mathrm{F}\big), \quad \text{for } k = 0, \dots, K-1.
\label{eq:update}
\end{equation}

\subsection{Unsupervised Objectives for Adaptation}

In the absence of ground-truth transcriptions during inference, we leverage unsupervised objectives based on the model’s own predictions. Let $H(p) = -\sum_{q \in \mathcal{Q}} p(q) \log p(q)$ denote the entropy of a probability distribution, and let $\mathrm{CE}(\tilde{y}, p) = -\log p(\tilde{y})$ denote the cross-entropy for a one-hot pseudo-label $\tilde{y}$. Our first objective, entropy minimization, encourages the model to produce confident predictions by reducing the entropy of its output distributions:
\begin{equation}
    \mathcal{L}_\mathrm{ent}(\mathcal{B}) = \frac{1}{B} \sum_{b=1}^{B} \sum_{t=1}^{T_b} H(p_{b,t}).
\label{eq:tent}
\end{equation}
Complementary to this, the pseudo-labeling objective generates self-training targets by selecting the most likely token at each step, $\tilde{y}_{b,t} = \arg\max_{q \in \mathcal{Q}} p_{b,t}(q)$, and minimizes the cross-entropy loss against these pseudo-labels:
\begin{equation}
    \mathcal{L}_\mathrm{pl}(\mathcal{B}) = \frac{1}{B} \sum_{b=1}^{B} \sum_{t=1}^{T_b} \mathrm{CE}(\tilde{y}_{b,t}, p_{b,t})
\label{eq:pl}
\end{equation}

\subsection{Confidence-Aware Filtering}

The core principle of confidence-aware filtering is to update the model using only those predictions in which it is highly confident, thereby focusing the adaptation on reliable signals.

We define the confidence of a prediction at time step $t$ for a sample $b$ as the maximum probability in its output distribution: $c_{b,t} = \max_{q \in \mathcal{Q}} p_{b,t}(q)$. A predefined confidence threshold $\tau$ is used to create a binary mask:
\begin{equation}
m_{b,t} = \mathbb{I}(c_{b,t} \geq \tau),
\label{eq:mask}
\end{equation}
where $\mathbb{I}(\cdot)$ is the indicator function. This mask $m_{b,t}$ equals 1 for confident predictions and 0 otherwise. It is applied to the loss computation, ensuring that only confident predictions contribute to parameter updates and guiding the model to learn from more reliable signals. We apply the confidence mask to each token-level loss and compute the final objective by averaging over only the confident tokens. If no tokens pass the threshold, the update is skipped.

\section{Experimental Setup}
\label{sec:experimental_setup}

For our experiments, we used Phi-4-Multimodal\cite{abouelenin2025phi}, a small open-source SLM with 5.6 billion parameters that accepts both audio and text inputs and generates text outputs. This model was selected for its strong performance relative to other models of comparable size. We evaluated the model across three task types. For automatic speech recognition (ASR), we used the LibriSpeech\cite{7178964} test-clean split. For speech translation (ST), we employed the CoVoST 2 dataset\cite{wang2020covost}, focusing on four bidirectional language pairs: German-to-English (DE-EN), English-to-German (EN-DE), Chinese-to-English (ZH-EN), and English-to-Chinese (EN-ZH). For closed-form question answering (QA), we used the AIR-Bench Foundation set, which includes closed-form QA tasks.

To assess robustness, we applied two types of corruption using noise from the WHAM! dataset. The first is additive noise in anechoic conditions, where randomly selected WHAM! noise was added at SNRs ranging from -5 to 5 dB. The second is a reverberant setting, where noise was added at SNRs of 10–20 dB and combined with simulated reverberation applied to both the clean signal and noise. Room impulse responses were generated using the pyroomacoustics toolkit\cite{scheibler2018pyroomacoustics}, simulating 400 rooms, each with 10 microphone arrays, one target source, and one noise source. Simulation parameters included absorption coefficients between 0.1 and 0.4, room dimensions ranging from $3{\times}3{\times}2$ to $10{\times}10{\times}5$ meters, and an image source method of order 6.

All experiments used a batch size of 5. For ASR and ST, we used the model’s default prompts: “Transcribe the audio clip into text” and “Translate the audio to \{lang\},” respectively. For the QA task, we modified the original prompt from “you may only choose A or B or C or D” to “answer with the full sentence” to encourage more complete generations. During test-time adaptation, the set of adaptable parameters $\theta_\mathrm{A}$ included all normalization layers, as recommended by the TENT framework\cite{wang2021tent}, and we further extended this set to include the convolutional subsampling layers in the audio encoder. In total, this yielded 2.58M adaptable parameters. We used a learning rate of $10^{-4}$ and experimented with 1, 3, and 5 adaptation steps. For confidence-aware filtering, we evaluated thresholds of 0.9, 0.8, 0.5, and 0.0, the last of which disables masking.

As there is limited research on TTA for SLMs, we followed the practice in \cite{lin22b_interspeech} and adopted dynamic pseudo-labeling as our baseline, which corresponds to our pseudo-labeling method with the confidence threshold set to 0. We also considered cascading a pre-trained denoising model before the SLM, but this consistently degraded performance under corruption and was therefore discarded.

\begin{table*}[t]
\centering
\small
\renewcommand{\arraystretch}{1.12}
\setlength{\tabcolsep}{5.2pt}
\caption{Test-time adaptation for \textbf{ASR}, \textbf{ST}, and \textbf{AIR-Bench (Foundation)}. 
ASR reports \textbf{WER}$\downarrow$; ST reports \textbf{BLEU}$\uparrow$ (ST cells show \emph{X$\to$EN\,/ EN$\to$X}, with X averaging DE and zh\_CN for X$\to$EN; EN$\to$X averages EN$\to$DE and EN$\to$zh\_CN); AIR-Bench reports \textbf{Accuracy}$\uparrow$. 
$^{\dagger}$ denotes the \emph{baseline: dynamic pseudo-labeling (PL) with no confidence masking}. 
Red = best $\Delta$; underlined = second best $\Delta$ within each task/direction.}
\label{tab:unified_asr_st_air_metrics}
\scalebox{0.88}{
\begin{tabular}{l l c l | c c c c | c c}
\toprule
\multicolumn{4}{c|}{\textbf{Setup}} & \multicolumn{4}{c|}{\textbf{Values at adaptation steps}} & \multicolumn{2}{c}{\textbf{Improvements}} \\
\textbf{Task} & \textbf{Dataset} & \textbf{Conf.} & \textbf{Objective} & \textbf{@0} & \textbf{@1} & \textbf{@3} & \textbf{@5} & $\boldsymbol{\Delta}$ & \textbf{Rel.} \\
\midrule
\multirow{8}{*}{ASR} & \multirow{8}{*}{\makecell[l]{LibriSpeech\\\footnotesize test-clean}}
& ✘            & PL$^{\dagger}$ & \multirow{8}{*}{5.83\%} & 5.75\% & 5.65\% & 5.56\% & 0.27\% & 4.63\% \\
& & ✘            & entropy       &                        & 5.47\% & 5.23\% & 5.49\% & 0.60\% & 10.29\% \\
& & $\tau{=}0.5$ & PL            &                        & 5.61\% & 5.47\% & 5.38\% & 0.45\% & 7.72\% \\
& & $\tau{=}0.5$ & entropy       &                        & 5.43\% & 5.27\% & 5.03\% & 0.80\% & 13.72\% \\
& & $\tau{=}0.8$ & PL            &                        & 5.61\% & 5.87\% & 5.16\% & 0.67\% & 11.49\% \\
& & $\tau{=}0.8$ & entropy       &                        & 5.54\% & 5.23\% & 4.99\% & \textcolor{red}{0.84\%} & \textcolor{red}{14.41\%} \\
& & $\tau{=}0.9$ & PL            &                        & 5.60\% & 5.36\% & 5.24\% & 0.59\% & 10.12\% \\
& & $\tau{=}0.9$ & entropy       &                        & 5.57\% & 5.25\% & 5.02\% & \underline{0.81\%} & \underline{13.89\%} \\
\midrule
\multirow{8}{*}{ST} & \multirow{8}{*}{\makecell[l]{CoVoST~2\\\footnotesize X$\to$EN\,/\\\footnotesize EN$\to$X}}
& ✘            & PL$^{\dagger}$ & \multirow{8}{*}{25.33 / 31.41} & 25.63 / 32.02 & 26.02 / 32.33 & 26.31 / 32.48 & 0.98 / 1.07 & - \\
& & ✘            & entropy       &                               & 25.96 / 32.20 & 26.69 / 32.80 & 26.98 / 33.38 & \textcolor{red}{1.66} / \textcolor{red}{1.97} & - \\
& & $\tau{=}0.5$ & PL            &                               & 25.67 / 31.86 & 25.99 / 32.48 & 26.46 / 32.75 & 1.13 / 1.33 & - \\
& & $\tau{=}0.5$ & entropy       &                               & 25.75 / 32.14 & 26.31 / 32.83 & 26.85 / 33.16 & \underline{1.52} / \underline{1.75} & - \\
& & $\tau{=}0.8$ & PL            &                               & 25.60 / 31.91 & 26.15 / 32.33 & 26.61 / 32.62 & 1.29 / 1.20 & - \\
& & $\tau{=}0.8$ & entropy       &                               & 25.67 / 31.98 & 26.11 / 32.67 & 26.53 / 32.72 & 1.21 / 1.31 & - \\
& & $\tau{=}0.9$ & PL            &                               & 25.68 / 31.82 & 26.05 / 32.21 & 26.24 / 32.45 & 0.91 / 1.04 & - \\
& & $\tau{=}0.9$ & entropy       &                               & 25.78 / 31.89 & 26.01 / 32.38 & 26.66 / 32.72 & 1.34 / 1.30 & - \\
\midrule
\multirow{8}{*}{QA} & \multirow{8}{*}{\makecell[l]{AIR\text{-}Bench\\\footnotesize Foundation}}
& ✘            & PL$^{\dagger}$ & \multirow{8}{*}{36.11\%} & 36.08\% & 36.17\% & 36.19\% & 0.08\% & 0.23\% \\
& & ✘            & entropy       &                          & 36.31\% & 36.62\% & 36.80\% & \textcolor{red}{0.69\%} & \textcolor{red}{1.91\%} \\
& & $\tau{=}0.5$ & PL            &                          & 36.00\% & 36.19\% & 36.25\% & 0.14\% & 0.39\% \\
& & $\tau{=}0.5$ & entropy       &                          & 36.19\% & 36.27\% & 36.43\% & 0.32\% & 0.88\% \\
& & $\tau{=}0.8$ & PL            &                          & 36.19\% & 36.23\% & 36.26\% & 0.15\% & 0.42\% \\
& & $\tau{=}0.8$ & entropy       &                          & 36.10\% & 36.45\% & 36.47\% & \underline{0.36\%} & \underline{0.99\%} \\
& & $\tau{=}0.9$ & PL            &                          & 36.11\% & 36.14\% & 36.11\% & 0.03\% & 0.08\% \\
& & $\tau{=}0.9$ & entropy       &                          & 36.13\% & 36.23\% & 36.37\% & 0.26\% & 0.72\% \\
\bottomrule
\end{tabular}}
\end{table*}

\begin{table}[t]
\centering
\small
\renewcommand{\arraystretch}{1.1}
\setlength{\tabcolsep}{6.5pt}
\caption{Reverberant corruption: baseline vs. best configuration per task. ST reports BLEU (X→EN / EN→X); ASR and AIR report absolute \% changes.}
\label{tab:reverb_small_summary}
\scalebox{0.9}{
\begin{tabular}{l c c c c}
\toprule
\textbf{Task} & \textbf{Conf.} & \textbf{Obj.} & \textbf{@0} & $\boldsymbol{\Delta}$ \\
\midrule
\multirow{2}{*}{ASR} 
  & ✘            & PL$^{\dagger}$ & \multirow{2}{*}{32.73\%}        & 4.01\% \\
  & ✘            & entropy        &                                 & 6.41\% \\
\midrule
\multirow{2}{*}{ST}  
  & ✘            & PL$^{\dagger}$ & \multirow{2}{*}{19.07 / 25.66}  & 1.21 / 1.45 \\
  & ✘            & entropy        &                                 & 2.27 / 2.71 \\
\midrule
\multirow{2}{*}{QA} 
  & ✘            & PL$^{\dagger}$ & \multirow{2}{*}{33.00\%}        & 0.18\% \\
  & ✘            & entropy        &                                 & 0.79\% \\
\bottomrule
\end{tabular}
}
\end{table}

\section{Results}
\label{sec:results}

As shown in Table \ref{tab:unified_asr_st_air_metrics}. For ASR, the best performance is achieved using entropy minimization with confidence-aware thresholding at $\tau{=}0.8$, resulting in a 0.84\% absolute WER reduction (14.41\% relative) compared to the non-adapted model, and outperforming the dynamic pseudo-labeling baseline ($\dagger$) by 0.57\% absolute and 9.78\% relative.
For speech translation, we evaluate both $X{\to}\mathrm{EN}$ and $\mathrm{EN}{\to}X$ directions. The strongest BLEU gains occur with entropy loss and no confidence masking ($\tau{=}0$), with improvements of +1.66 and +1.97 BLEU, respectively, over the non-adapted model. These outperform the pseudo-labeling baseline by +0.68 and +0.9 BLEU.
For QA tasks in AIR-Bench, entropy loss with $\tau{=}0$ again performs best, improving accuracy by +0.69\% (relative +1.91\%) and exceeding the baseline by +0.61\% absolute (relative +1.68\%).

Across nearly all matched settings, the entropy objective outperforms pseudo-labeling.
The preferred confidence threshold differs by task: ASR peaks at a higher threshold ($\tau{=}0.8$), while ST and AIR-Bench peak with no masking ($\tau{=}0$).
We hypothesize that ASR’s stronger base accuracy leaves enough high-confidence tokens even at higher $\tau$, which suppresses noisy updates and stabilizes adaptation; in contrast, for ST and AIR-Bench, a high $\tau$ filters out too many updates. With $\tau{=}0$, entropy minimization can leverage the full predictive distribution, providing useful gradients even when some predictions are imperfect.

We further evaluate in a more challenging setting with noisy reverberation. Table \ref{tab:reverb_small_summary} contrasts the baseline performance with our best-performing configuration. Across all tasks, the entropy objective without confidence masking ($\tau{=}0$) remains the best configuration and consistently surpasses the pseudo-labeling baseline: ASR gains 6.41\% WER reduction versus 4.01\% for the baseline; ST improves by 2.27 / 2.71 BLEU versus 1.21 / 1.45 BLEU; AIR-Bench gains 0.79\% accuracy versus 0.18\%. The absolute improvements are larger than in the anechoic condition, indicating that test-time adaptation is especially beneficial under stronger shifts. Notably, for ASR the optimal threshold moves from $\tau{=}0.8$ (anechoic) to no masking under reverberation, aligning with our hypothesis that harsher corruptions reduce the share of high-confidence tokens; relaxing the mask increases the number of effective updates and avoids over-filtering.
\section{Conclusion}
We proposed SLM-TTA, a parameter-efficient test-time adaptation framework for generative spoken language models with interleaved audio–text prompts. Our method integrates entropy minimization and pseudo-labeling with confidence-aware token masking, and performs episodic adaptation on a small subset of model parameters without requiring source data or labels. Experiments across ASR, speech translation, and closed-set QA tasks demonstrate consistent robustness improvements under both anechoic and reverberant conditions, with minimal computational overhead. Future directions include extending SLM-TTA to a wider range of model architectures, incorporating open-ended chat-style tasks, and addressing more diverse real-world acoustic distortions.

\bibliographystyle{IEEEbib}
\bibliography{strings,refs}

\end{document}